\documentclass[twocolumn,showpacs,preprintnumbers,amsmath,amssymb,aps,pre]{revtex4}

\usepackage{graphicx}
\usepackage{dcolumn}
\usepackage{bm}
\usepackage{amsmath}
\usepackage{hyperref}

\begin{document}

\title{Coherent wave transmission in quasi-one-dimensional systems with L\'evy  disorder}
\author{Ilias Amanatidis$^1$, Ioannis Kleftogiannis$^{2,\dagger}$}
\affiliation{$^1$Center for Theoretical Physics of Complex Systems, Institute for Basic Science (IBS), Daejeon 34051, Republic of Korea}
\affiliation{$^2$Department of Physics, National Taiwan University, Taipei 10617, Taiwan}
\author{Fernando Falceto, V\'ictor A. Gopar}
\affiliation{Departamento de F\'isica Te\'orica and BIFI, Universidad de Zaragoza, Pedro Cerbuna 12, E-50009, Zaragoza, Spain}

\begin{abstract}

We study the random fluctuations of the transmission in disordered  quasi-one-dimensional  systems such as disordered
waveguides and/or  quantum wires whose random configurations of disorder are characterized by density distributions with a  long tail   known as  L\'evy
distributions. The presence of
L\'evy disorder leads to  large fluctuations of the transmission and anomalous localization, in relation to the standard exponential
localization (Anderson localization).
We calculate the complete distribution of the transmission fluctuations for different number of
transmission channels in the presence and absence of time-reversal symmetry. Significant  differences in the  transmission statistics between disordered systems
with Anderson and anomalous
localizations are revealed. The theoretical predictions are independently confirmed by  tight binding numerical simulations.

\end{abstract}


\maketitle

\section{Introduction}

Coherent wave-interference phenomena have been experimentally and theoretically investigated in different complex systems such
as disordered waveguides, photonic crystals, cold atoms, and disordered quantum wires. One of the most celebrated effects of waves in
random media is the Anderson localization: an exponential decay in  space  produced by destructive interference. The phenomenon of Anderson
localization was originally predicted  for electrons [\onlinecite{Anderson}],   but being fundamentally a wave phenomenon, Anderson localization
has been also observed  in electromagnetic and acoustic experiments,    Bose-Einstein condensates and
entangled photons, see for instance [\onlinecite{50Anderson, Fifty_years,Anderson_photons, Bose-Einstein,Chabanov,Shi,Penha,Yamilov}].

Wave scattering in complex  media has been widely  investigated  within different approaches. In particular,
the so-called Dorokhov-Mello-Pereyra-Kumar (DMPK) equation [\onlinecite{Mello_book}]  has been successfully applied to study
several  statistical properties of
electronic  transport through disordered quantum  wires, as well as classical waves in disordered structures. The DMPK equation is a diffusion
equation that describes the evolution of the probability density of the transmission eigenvalues  as a function of the length of the
system. Remarkably,
within this approach, the statistical properties of the transmission depend only of a few physical  parameters of the system
such as the localization length and the presence (or not) of  time reversal symmetry, i.e.,  all other details of the system are   irrelevant for  the
statistical description of the transport.

The diffusion approach to wave transport (DMPK equation)  has been applied to study different statistical properties of
 functions of the transmission [\onlinecite{Mello_book,Beenakker-review}] in systems where quantum wave functions (electrons) or classical electromagnetic waves  are
 exponentially localized in space (Anderson localization).

The presence of  disorder, however, does not
necessarily  lead to the standard Anderson localization. Actually, nonstandard localization can be produced by different means in random media.
For instance, it has been shown that electrons in disordered quantum wires at the band center [\onlinecite{Soukoulis, Evangelou,We}]  and
armchair graphene disordered  nanoribbons [\onlinecite{nanoribbons}] are anomalously localized.
In particular, it has been experimentally and theoretically
shown that random  configurations  of the disorder characterized by  probability densities  with a power-law tail (also known as L\'evy  distributions) produce
anomalous localization, i.e., waves are weaker localized in space in relation to the standard Anderson localization. For experimental realizations of L\'evy disorder, see for instance
Refs. [\onlinecite{Barthelemy, Antonio_prl}].  Those
works, however, have been restricted to  one-dimensional systems or structures where  only a single transmission eigenvalue or transmission channel is
relevant. It is also worth to mention
that L\'evy-type distributions have been used to study different problems in a wide range of science disciplines [\onlinecite{Michael, Lambert, stock,Hideo,fishes,Mercadier}].

In general, however,  the  transmission through a system is given by the contribution of several transverse modes or transmission channels [Eq. (\ref{transmission})]. Therefore,
it is highly desirable going beyond the single channel case and
consider  the possibility that several transmission channels contribute to the transport, which is also  a less restrictive condition from an  experimental point of view.
 Additionally, the multichannel case allows to study the effect of the absence (or presence) of time-reversal symmetry
in  Levy  disordered systems.

With the above motivation, in this work  we extend  the diffusion approach to consider the case of anomalous
localization in quasi-one dimensional disordered systems, where several transmission channels play a role, i.e., we study the statistical
properties of the transmission of  waves that are anomalously localized in relation to  those with
 standard Anderson localization.  In order to induce anomalous localization, we shall consider that the random configurations of the scatterers in a quasi 1D disordered
system follow a distribution with a long tail:  if $x$ is a random variable with probability density $\rho(x)$, then for large $x$,
$\rho(x) \sim 1/x^{1+\alpha}$ with $0< \alpha <2$. This kind of distributions is also known as L\'evy type distributions or
$\alpha$-stable distributions [\onlinecite{Levy,Kolmogorov,Uchaikin}].  We notice that for $0 < \alpha <1$, the first  moment of $\rho(x)$ diverges.
In this work we shall consider the range $0 < \alpha  < 1$, where  effects of Levy disorder are strong, as we  shall show [\onlinecite{largervalues}].

The present work is an extension of the
one-dimensional case  studied in Ref. [\onlinecite{Falceto}] to the multichannel case. This extension allows to investigate the transport properties of the
transmission  under physical conditions that cannot be considered in the 1D case such as the effects of breaking the time-reversal symmetry of the system.
 All our theoretical
predictions are independently confirmed by  numerical simulations of quasi-one dimensional disordered systems  using a tight-binding model.

The remainder  of this  paper is organized as follows.
For the sake of completeness, the  next section  is devoted to the problem of transport in 1D disordered systems or a single transmission channel. Both standard and
anomalous localizations are studied and some previous results of  Ref. [\onlinecite{Falceto}] are reproduced. In section \ref{multichannel_section}, we extend our results to the
multichannel transmission case. We first  briefly introduce some elements of the DMPK equation whose solution gives the join probability density function of $N$ transmission
channels, which  is used  later to calculate the transmission distribution in presence of L\'evy disorder for systems supporting
an arbitrary number  of    channels. Within the same  section \ref{multichannel_section},  several examples of the transmission distribution for L\'evy disordered systems are
shown for systems which  preserve or break time-reversal symmetry. Finally in section \ref{conclusions_section}, we give a summary and conclusions of our work.

\section{Single transmission channel}\label{single}

Disordered systems with L\'evy-type disorder [\onlinecite{incoherent}] and  a single transmission channel were
studied  in Ref. [\onlinecite{Falceto}]. Here, we briefly present this case for the sake of completeness and since it is used to derive the length dependence
of the multichannel transmission, as we show below.

Thus, following Ref. [\onlinecite{Falceto}], we consider an one-dimensional disordered system  with scatterers randomly placed along its
length $L$. The key ingredient in this model is that the random distance between the scatterers follows a distribution with a long tail. To obtain the statistical
properties of the transmission in the presence of L\'evy-type disorder, we extend the results of random-matrix calculations for standard disordered systems.

\subsection{Standard localization}\label{standard}

As previously mentioned,
the scaling approach to localization and random-matrix theory  has been extensively developed in the past
[\onlinecite{Anderson_1,Mello_book, Beenakker-review}] and  applied to describe  the statistical
properties of transport in standard disordered media, i.e., systems whose disorder models involve  distributions with finite mean values.
Within the scaling theory framework, a diffusion-type equation for the probability distribution of the transmission $T$ was derived and conveniently written
in terms of the variable $\lambda$ as [\onlinecite{Melnikov}]:
\begin{equation}
\label{1Ddiffusion}
l \frac{p_s(\lambda)}{\partial L}=\frac{\partial}{\partial \lambda}\left[\lambda\left(\lambda+1 \right) \frac{\partial p_s(\lambda)}{\partial \lambda} \right]
\end{equation}
where $\lambda=1/(1+T)$. The solution of  Eq. (\ref{1Ddiffusion}) can be written as [\onlinecite{Abrikosov}]
\begin{equation}
\label{pofg}
p_s(T)=\frac{s^{-\frac{3}{2}}}{\sqrt{2\pi}} \frac{{\rm
e}^{-\frac{s}{4}}}{T^2}\int_{y_0}^{\infty}dy\frac{y{\rm
e}^{-\frac{y^2}{4s}}} {\sqrt{\cosh{y}+1-2/T}},
\end{equation}
where $y_0={\rm arcosh}{(2/T-1)}$ and $s=L/l$. We point out that  the distribution of the transmission is determined  by a single
microscopic property of the system: the mean free path $l$, in $s=L/l$.

From the distribution $p_s(T)$ we can obtain any average value of the transmission. In particular,  an exponential decay
of the average transmission with the length is found:
\begin{equation}
\label{averT}
 \langle T \rangle \propto \exp{(-L/2l)} ,
\end{equation}
while the average of the logarithm of the transmission is given by
\begin{equation}
\label{logT}
 \langle - \ln T \rangle = L/l .
\end{equation}
We notice that the mean free path can thus be obtained from $\langle - \ln T \rangle$.
For later purposes, at this point we also remark that  $\langle - \ln T \rangle$ is a linear function of $L$.

Let us illustrate the above results [Eqs. (\ref{pofg})-(\ref{logT})].  This will be useful
for  contrasting the effects of anomalous localization due to the presence of L\'evy-type disorder  in the next section.

\begin{figure}
\includegraphics[width=0.9\columnwidth]{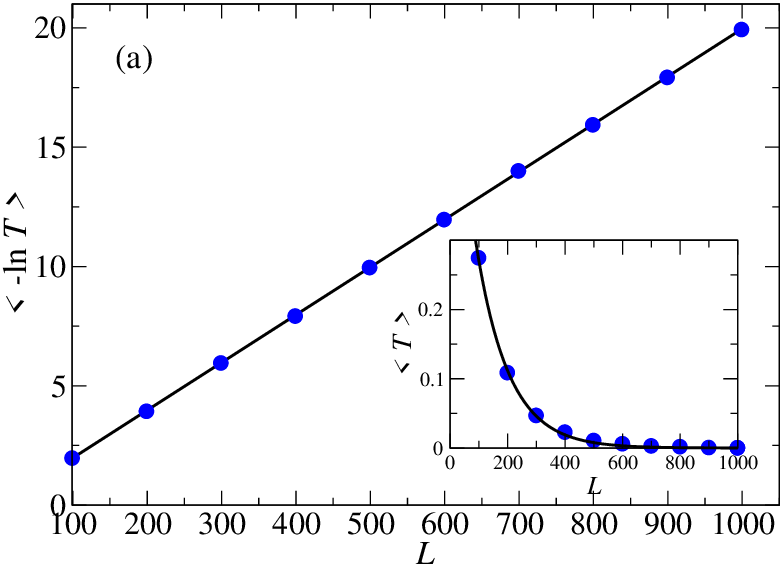}
\includegraphics[width=0.94\columnwidth]{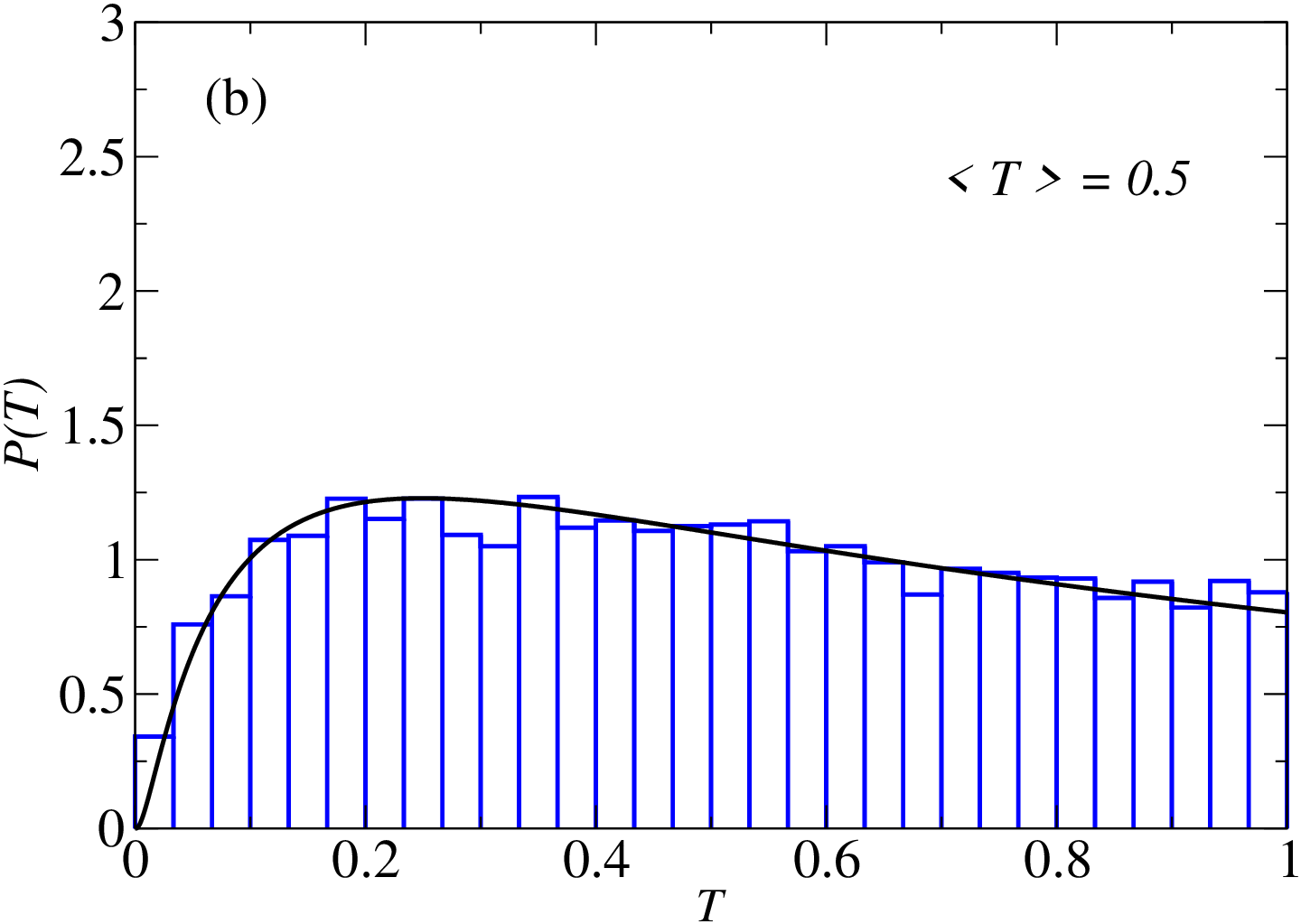}
\caption{(Color online) Transmission results for 1D systems with standard localization. (a) Linear (main frame) and exponential (inset)  dependence of $\langle \ln T \rangle$ and
$\langle T \rangle$, respectively, with
 the system length $L$. The  solid lines correspond to the theoretical results, while the solid dots
 are obtained from the numerical simulations. (b) The complete distribution  of the transmission $P(T)$ for a standard
 disordered system  with average $\langle T \rangle=0.5$.  The solid line correspond to the theoretical distribution, while the histogram is extracted from the numerical simulations.}
\label{fig_1}
\end{figure}

In Fig. \ref{fig_1} (a), it is shown the linear (main frame) and exponential decay (inset) behavior of the averages
$\langle - \ln T \rangle$ and $\langle  T \rangle$  in Eqs. (\ref{averT}) and (\ref{logT}), respectively.  The results of
numerical simulations (dots), using a tight-binding model (see Appendix),  are in agreement with the theoretical ones (solid lines). In Fig. \ref{fig_1} (b),
we show the transmission distribution (solid line) as given in Eq. (\ref{pofg}) for
$s=0.93$, while the histogram corresponds to the transmission distribution obtained from the  tight-binding simulations (see Appendix). Thus, we can observe
that Eq. (\ref{pofg}) and the numerical simulations are in good agreement.

\subsection{Anomalous localization}\label{anomalous_section}

We now introduce a  L\'evy-type model for the disorder  that leads to anomalous localization. Following
Ref. [\onlinecite{Falceto}], we consider that  $\nu$ scatterers are randomly distributed in a system of length $L$ and assume that their separation
$x$  follows a probability density with a long tail:
\begin{equation}
 \rho(x) \sim \frac{c}{x^{1+\alpha}},
\end{equation}
for large $x$. Here $c$ is a constant.  As we already mentioned, the first moment of $\rho(x)$ diverges which  implies that the mean free path $l$ diverges. We recall that $l$
governs the statistics of the transmission in
the standard localization problem, as pointed out in the previous subsection. Therefore, we might expect a nonstandard behavior on the transmission statistics in the presence of
L\'evy disorder.

Let us define the probability density $\Pi_L(\nu)$ of the number of scatterers in a system
of length $L$. It has been shown [\onlinecite{Falceto}] that $\Pi_L(\nu)$ is given in terms of the probability density distribution $q_{\alpha,c}(x)$ of
the L\'evy distribution as
\begin{equation}
\label{pi_alpha<1}
 \Pi_{L}(\nu)=\frac2\alpha\frac{L}{(2\nu)^{\frac{1+\alpha}{\alpha}}}
q_{\alpha,c}({L}/{(2\nu)^{1/\alpha}}) ,
\end{equation}
for $0 < \alpha <1 $, in the limit $L\gg c^{1/\alpha}$. We remark that $q_{\alpha,c}(x)$ has a power-law tail [\onlinecite{qalpha_fourier}]:
$q_{\alpha,c}(x) \sim c/x^{1+\alpha}$ for large values of  $x$.
\begin{figure}
\includegraphics[width=0.9\columnwidth]{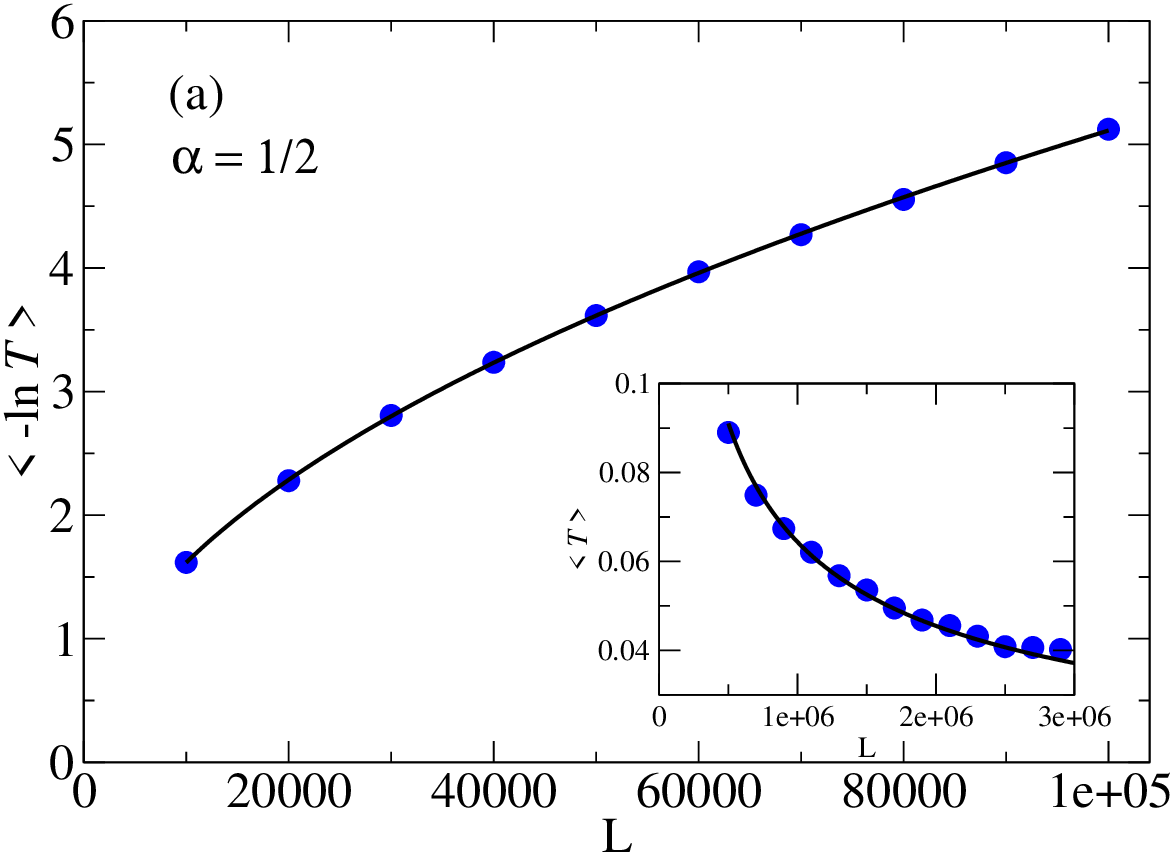}
\includegraphics[width=0.94\columnwidth]{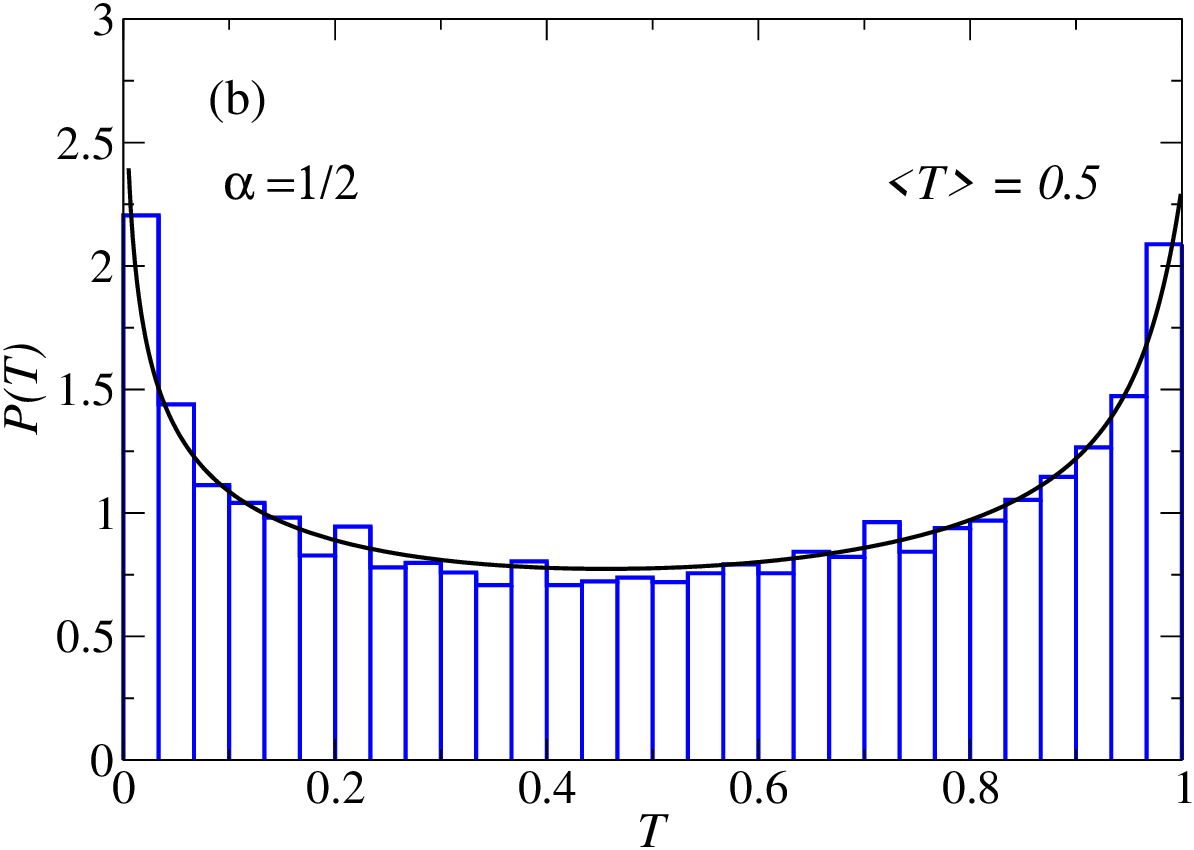}
\caption{(Color online) Transmission results for systems with L\'evy disorder ($\alpha=1/2$). (a) Power law dependence of $\langle \ln T \rangle$ (main frame)  and $\langle T \rangle$  (inset) for a 1D system with L\'evy disorder  with $\alpha=1/2$. See Eqs.
(\ref{lnTaver_alpha})  and (\ref{Taver_alpha}). Solid dots correspond to the numerical simulation results. (b) The complete distribution $P(T)$  for a L\'evy disordered
system with $\alpha=1/2$ and $\langle T \rangle=0.5$. The solid line is obtained from Eq. (\ref{pofG_xi}), while the histogram is obtained from the numerical simulations.}
\label{fig_2}
\end{figure}

We now introduce the average values  $\langle \ln T\rangle_\nu$ and $\langle \ln T\rangle_L$  for systems with a fixed number of scatterers
$\nu$ and fixed length $L$, respectively.
From the standard scaling theory of localization summarized in previous subsection,
$\langle -\ln T\rangle_\nu$ is proportional to $\nu$: $\langle -\ln T\rangle_\nu
= a \nu$, $a$ being a constant [\onlinecite{Mello_groups}]. Hence,  we have that
\begin{eqnarray}
\label{lngofl}
\langle -\ln T\rangle_L&=&\int_0^\infty  \langle -\ln T\rangle_\nu  \Pi_{L}(\nu) {\rm d}\nu  \\
&=&\int_0^\infty a \nu \frac2\alpha\frac{L}{(2\nu)^{\frac{1+\alpha}{\alpha}}}
q_{\alpha,c}({L}/{(2\nu)^{1/\alpha}}) {\rm d}\nu ,
\end{eqnarray}
where we have substituted  Eq. (\ref{pi_alpha<1}). Using the scaling property of
the L\'evy distributions: $c^{1/\alpha}q_{\alpha,c}(c^{1/\alpha}x)=q_{\alpha,1}(x)$,  and introducing the
 variable $ z=L/(2c\nu)^{1/\alpha}, $ we  obtain
\begin{equation}
\label{lngofl_a}
\langle -\ln T\rangle_L = L^\alpha  \frac{a}{c} \frac12\int_0^\infty z^{-\alpha}q_{\alpha,1}(z){\rm d}z
= L^\alpha \frac{a}{c} I_\alpha,
\end{equation}
where $I_\alpha =(1/2)\int_0^\infty z^{-\alpha}q_{\alpha,1}(z){\rm d}z=\cos(\pi\alpha/2)/2\Gamma(1+\alpha)$ [\onlinecite{Engin}], here $\Gamma$ is the Gamma function.
We point out that Eq. (\ref{lngofl_a}) shows a nonstandard behavior:
\begin{equation}
\label{lnTaver_alpha}
 \langle -\ln T\rangle_L\propto L^\alpha ,
\end{equation}
i.e., $\langle -\ln T\rangle_L $ is a power function of $L$,  in contrast
to the linear behavior with $L$ expected in the usual scaling theory [see Eq. (\ref{logT})].
Similarly, the average of the
transmission decays with the length as
\begin{equation}
\label{Taver_alpha}
 \langle T \rangle_L \sim 1/L^\alpha ,
\end{equation}
which is also in contrast to the expected exponential decay in 1D [see Eq. (\ref{averT})].

In Figs. \ref{fig_2}(a) and \ref{fig_2}(b) we have verified the above
results [Eqs. (\ref{Taver_alpha}), (\ref{lnTaver_alpha})] for $\alpha=1/2$  by comparing with numerical simulations.  As it was predicted, $\langle \ln T \rangle$ has a
 power-law behavior with $L$, in this case with power $\alpha =1/2$, while $\langle T \rangle$ decays as $L^{-1/2}$.

We now calculate the complete distribution of the transmission $P_\mu (T)$ for fixed length $L$  in the presence
of L\'evy disorder, here  $\langle -\ln T \rangle_L=\mu$.

The  distribution  $P_\mu (T)$ can be obtained from $p_s(T)$,  Eq. (\ref{pofg}), using  that
in the standard diffusion approach the parameter $s$  is proportional
to the number of scatterers $\nu$, i.e., $s=a \nu$, $a$ being a constant. Thus, we introduce the information of
the L\'evy disorder through the distribution $\Pi_{L}(\nu)$ in Eq. (\ref{pi_alpha<1}) to obtain
that the probability density of the transmission $P_\mu (T)$ is given by
\begin{eqnarray}
P_\mu(T)
&=&
\int_0^\infty
p_{a\nu}(T)
\Pi_{L}(\nu)
{\rm d}\nu.
\end{eqnarray}
Using Eqs. (\ref{pi_alpha<1}) and (\ref{lngofl_a}) as well as the
scaling properties of the L\'evy distributions, we finally have that
\begin{eqnarray}
\label{pofG_xi}
P_\mu(T)=\int_0^\infty p_{s(\alpha,\mu,z)}(T) q_{\alpha,1}(z){\rm d}z ,
\end{eqnarray}
where we have defined
$
s(\alpha,\mu,z)={\mu}/(2{z^\alpha I_\alpha)}.
$
We remark  that the distribution  $P_\mu(T) $ in Eq. (\ref{pofG_xi}) depends
only on two parameters $\langle -\ln T \rangle_L =\mu$ and $\alpha$, i.e., other details of
the disorder configurations are irrelevant.

As an example, in Fig. \ref{fig_2} (b)
we show the complete transmission distribution for a disordered system with $\langle T \rangle =0.5$ and $\alpha=1/2$. It is interesting to compare the distribution of
the transmission for  both standard and
anomalous localizations,  Figs. \ref{fig_1}b and \ref{fig_2}b, respectively. Notice  that both distributions are obtained for
disordered systems with $\langle T \rangle =0.5$, however, the shape of the distributions are totally different. In particular, in the case of
anomalous localization, the transmission distribution show two peaks at $T=0$ and $T=1$, which is a consequence of the stronger random
fluctuations of the transmissions in the presence of L\'evy-type disorder.

\section{Multichannel transmission}\label{multichannel_section}

In the previous section,  we have considered the simplest case of 1D disordered systems where only a single transmission channel plays a role.
We now extend our analysis to a more general case where the total transmission is given by the contribution of several transmission
channels.  Additionally,
by considering the multichannel case we can  study the effects of
the presence, or absence,  of time-reversal symmetry. We shall present  particular cases of two and three transmission channels to
illustrate our results.  Similarly to the previous section \ref{single}, we first give a summary of the  random-matrix theory for standard disorder and later we
extend the  results to consider L\'evy-type disorder.

\subsection{Standard localization}

Let us consider a disordered system  whose length $L$ is much larger than its width, i.e. a quasi-one dimensional system.  With this geometry, one  can
neglect  diffusion in the transverse direction. Assuming that
the system supports $N$  transverse modes, or  channels, the total transmission $T$ is   given by
\begin{equation}
\label{transmission}
 T=\sum_{n=1}^N \tau_n ,
\end{equation}
where $\tau_n$ are the eigenvalues of the  product $tt^\dagger$, being $t$ the matrix of the transmission amplitudes of a quasi-1D
disordered system. Within the diffusion approach [\onlinecite{dmpk}], the transmission eigenvalues $\tau_n$ are random variables whose joint probability
distribution  function $p(\tau)$  evolves with the system length $L$
according  to a Fokker-Planck equation, or DMPK equation,  as [\onlinecite{dmpk,Mello_book}]
\begin{eqnarray}
\label{dmpk}
l \frac{\partial p (\lambda)}{\partial L}=\frac{2}{\beta N+2-\beta}\frac{1}{J(\lambda)}\sum_i^N&& \frac{\partial}{\partial \lambda_i}
\left[ \lambda_i \left( 1+\lambda_i \right) J(\lambda) \right.   \nonumber \\ &&
\left. \times \frac{\partial}{\partial \lambda_i}\frac{\partial p(\lambda)}{J(\lambda)}  \right] ,
\end{eqnarray}
where  $\lambda_i=(1-\tau_i)/\tau_i$, while $l$ is the mean free path. The Jacobian
$J(\lambda)$ is given by the product  $ J(\lambda)= \prod_{i<j}^N|\lambda_i-\lambda_j | ^\beta$. The value of the parameter $\beta$ depends on
the absence ($\beta=2$) or presence ($\beta=1$) of time reversal symmetry.  The above diffusion equation [Eq. (\ref{dmpk})] is a generalization
of the single channel case in Eq. (\ref{1Ddiffusion}).
We also notice that the mean free path $l$ is the only microscopic information that enters into the diffusion equation, as in the
single-transmission channel problem in the previous section.

On the other hand, an analytical expression for the solution of the DMPK, Eq. (\ref{dmpk}),  for both unitary ($\beta=1$) and orthogonal ($\beta=2$) symmetries
has been obtained in the metallic  and insulating   regimes [\onlinecite{Rejaei}], which also has been useful to study statistical properties of the transmission in the
metal-insulating crossover regime [\onlinecite{crossover, Muttalib}]. This solution can be written as:
\begin{equation}
\label{plambda}
 p^{(\beta)}_s(\lambda)=\frac{1}{Z}\exp{(-\beta H(\lambda))} ,
\end{equation}
where $Z=\int \exp{(-\beta H(\lambda))} \Pi_i d\lambda_i$ and
\begin{equation}
 H(\lambda)=\sum_{i<j}^N u(\lambda_i,\lambda_j)+\sum_i^N V(\lambda_i)
\end{equation}
The functions $u(\lambda_i,\lambda_j)$ and  $V(\lambda_i)$ are more conveniently written in terms of the variables $x_i$, where $\lambda_i= \sinh^2 x_i$ as:
\begin{eqnarray}
 u(x_i,x_j)&=&-\frac{1}{2}\left[ \ln \left| \sinh^2 x_i -\sinh^2 x_j \right|-  \ln \left| x^2_i - x^2_j \right|   \right] ,  \nonumber \\
 V(x_i)&=&\frac{l(\beta N+2-\beta)}{2L\beta}x^2_i - \frac{1}{2 \beta}\ln\left| x_i \sinh2x_i\right|
\end{eqnarray}

Therefore, using the joint probability distribution given in Eq. (\ref{plambda}),
the distribution of the transmission is given by  the average
\begin{equation}
\label{pofG_multichannel}
p_{s}^{(\beta)}(T)=\left\langle \delta \left( T-\sum_i^N \frac{1}{1+\lambda_i}\right) \right\rangle ,
\end{equation}
where, as previously defined, $s$ is the length of the system in units of the mean free path ($s=L/l$) and
the brackets denote the average performed with the join probability distribution $p_s^{(\beta)}(\lambda)$, Eq. (\ref{plambda}).

 As we have mentioned, the above diffusion
approach have been  successfully verified in a number of numerical and experimental works where Anderson localization is present [\onlinecite{Mello_book,Beenakker-review}].

With the above results, we are now ready to introduce L\'evy-type disorder in a  multichannel disordered media.

\subsection{Anomalous localization}\label{anomalous_multichannel}

Let us assume the presence of L\'evy-type disorder, as described in the previous Section \ref{anomalous_section}, in a multichannel system of length $L$. In addition to the
interest in studying the transmission properties of L\'evy-type disorder media supporting many channels, the multichannel problem adds the possibility of
studying  the effects  of breaking the  time-reversal symmetry of the system, characterized by the parameter $\beta$. We shall  consider the cases of $\beta=1$
(preserved time-reversal symmetry) and $\beta=2$ (broken time-reversal symmetry).

The distribution of the transmission for L\'evy-type disordered systems in the multichannel case can be obtained following the steps of the one channel
case, Section II B.  Although the  generalization  to the multichannel case is straightforward, the calculations
are more involved and no simple  analytical relations have been obtained.

As in   the one channel case, the transmission distribution for multichannel L\'evy  disordered systems can be obtained once
the probability density of the number of scatterers in a system of fixed length $L$ is known, assuming the separation between scatterers
follows a L\'evy-type distribution. This probability density was already obtained in Section II B, Eq. (\ref{pi_alpha<1}). Therefore, with the
knowledge of transmission distribution from the standard random-matrix theory given by Eq. (\ref{pofG_multichannel}), we write the
density probability distribution of the transmission for L\'evy-type disorder as
\begin{eqnarray}
\label{pofG_mu}
P_{\mu}^{(\beta)}(T)=\int_0^\infty p_{s(\alpha,\mu,N,z)}^{(\beta)}(T) q_{\alpha,1}(z){\rm d}z ,
\end{eqnarray}
where the distribution $p^{(\beta)}_{s(\alpha,\mu,N, z)}(T)$ is given in Eq. (\ref{pofG_multichannel}) with $s$ replaced by a function
$s(\alpha,\mu,N, z)$ and $\mu =\langle \ln T \rangle_L$.  For the single transmission channel, we have given  an expression for $s(\alpha,\mu,N=1, z)$ in
terms of the
average $\langle \ln T \rangle_L$ since $s=\langle \ln T \rangle$ in the case of standard disorder.
In the multichannel case, however, we cannot derive  an analytical  expression for $s(\alpha,\mu,N, z)$ since  there is no a  general expression
between $s$ and  $\langle \ln T \rangle_L$ for arbitrary number of  channels. For $N \gg 1$, however,
$\langle \ln T \rangle \approx  L/(\beta N l)=s/(\beta N)$. Thus, in this limit, we can write
$s(\alpha,\mu,N,z)=(\beta N){\mu}/(2{z^\alpha I_\alpha)}$.  To overcome this problem for arbitrary number of channels, we consider that the function $s(\alpha,\mu,N, z)$ is of the form  $ b/z^\alpha$, where $b$ is a
constant whose value  is fixed to that one that reproduce the numerical value of the average $\langle T \rangle_L$, or
equivalently $\langle \ln T \rangle_L$.

We thus now present  several   examples of the transmission distribution as given by Eq. (\ref{pofG_mu}) for $N=2$ and 3 transmission channels
and different values of the power decay $\alpha$,
in the presence and absence of time reversal symmetry. The theoretical results are obtained by numerical integration of Eq. (\ref{pofG_mu}). In all  cases, our results are
independently verified  by
tight-binding numerical simulations. Additionally, in order to contrast and compare the transmission statistics of standard and L\'evy disordered systems,   in the first panel  of  the
following  Figs. \ref{fig_3}-\ref{fig_6} we include
the transmission distribution expected for  the cases of standard disordered systems.

\begin{figure}
\includegraphics[width=\columnwidth]{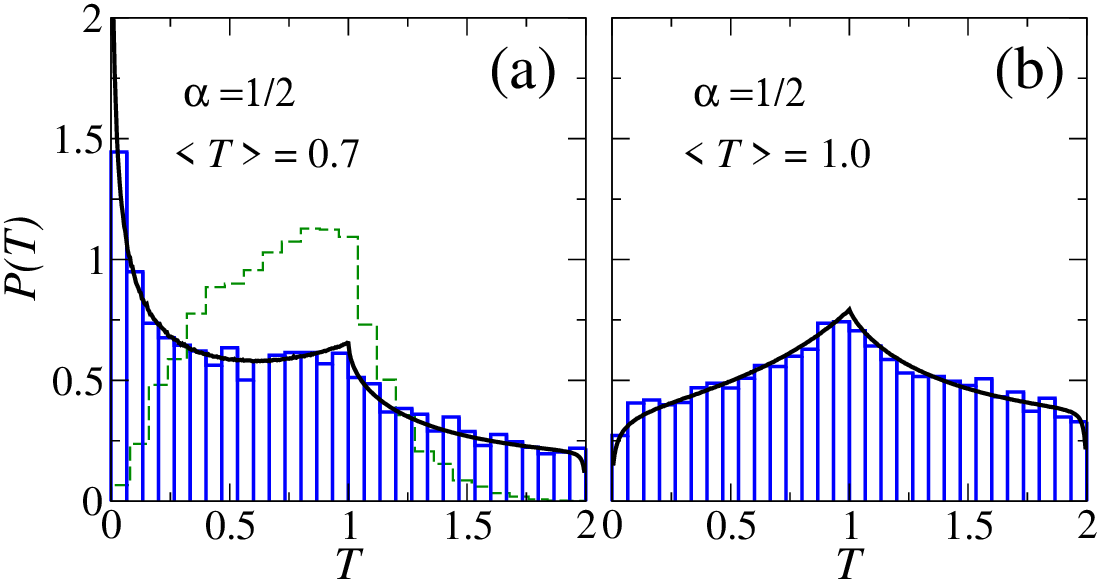}
\includegraphics[width=\columnwidth]{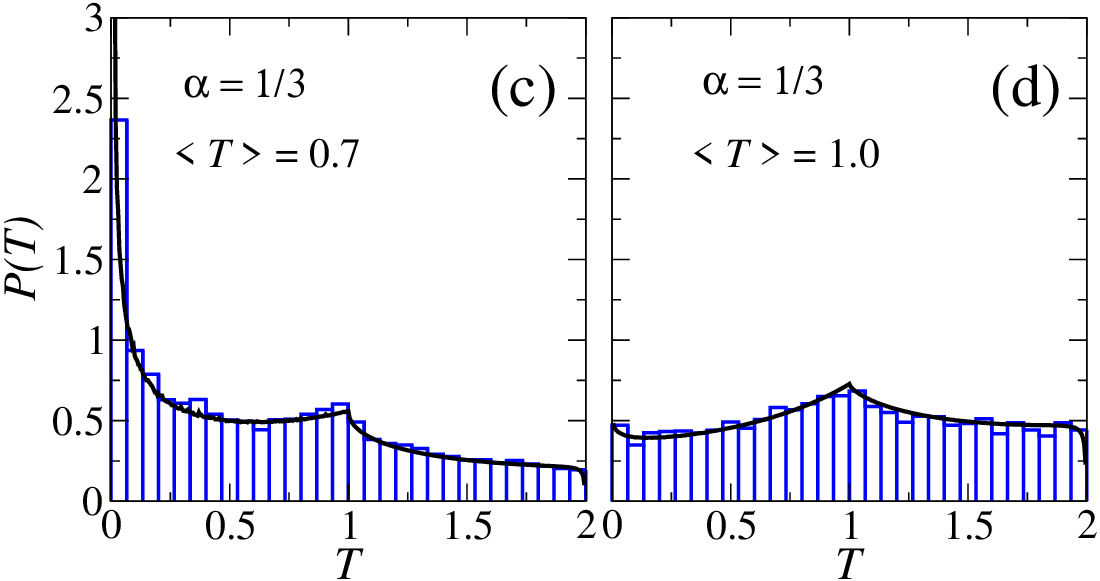}
\caption{(Color online) Transmission distribution for $N=2$ transmission channels with L\'evy disorder characterized by  $\alpha=1/2$ and 1/3, upper and lower
panels, respectively. The solid lines are obtained according to the theory described in the main text, while the histograms are extracted from the tight binding numerical simulations.
The values of the standard deviation $\delta T$ and the constant $b$ in $s(\alpha,\mu,N, z)$ for each panel are: (a) $\delta T=0.54$, $b=2.5$, (b) $\delta T=0.52$, $b=1.35$, (c) $\delta T=0.56$,
$b=2.6$, (d) $\delta T=0.54$, $b=1.2$. A good agreement between theory and numerical simulation can be seen in all panels.   For comparison with L\'evy disordered systems, in
panel (a)   it is shown $P(T)$  (green dashed-line histogram)  for systems with standard disorder and  ensemble average $\langle T \rangle =0.7$. }
\label{fig_3}
\end{figure}

\subsubsection{Preserved Time-reversal symmetry}

We first assume  that  time-reversal symmetry is present in the  system, i.e., we consider the symmetry class $\beta=1$ and,
as it was previously mentioned, we shall concentrate in the cases of $N=2$ and 3  channels.

The distribution of the transmission for 2  channels is shown in Fig. \ref{fig_3} for two  different values of the average $\langle T \rangle$  and disorder
configurations characterized by the decay power $\alpha=1/2$ and 1/3.

The histograms (blue solid line) in Fig. \ref{fig_3} are obtained by  tight-binding numerical  simulations (see Appendix) by  collecting the transmission data
from 10000  disorder configurations, while the theoretical predictions (black solid lines) are calculated according to
Eq. (\ref{pofG_mu}) with $p_{s(\alpha,\mu,N, z)}(T)$ given by  Eq. (\ref{pofG_multichannel}) with $N=2$ and $\beta=1$.

The distributions in Figs. \ref{fig_3}(a) and \ref{fig_3}(b) correspond to the case of $\alpha=1/2$ with
average transmission $\langle T \rangle = 0.7$ and 1.0, respectively. Similarly, Figs. \ref{fig_3}(c) and \ref{fig_3}(d) show the transmission distribution
for $\alpha=1/3$. We can observe a good agreement between  theoretical  (solid lines) and  numerical simulation results (histograms)   in all
cases.

It is expected that the fluctuations of the transmission become large as  the power exponent $\alpha$ decreases.
This implies  that the transmission distributions for $\alpha=1/3$ are  wider than those with $\alpha=1/2$. Effectively, for a fixed value of the $\langle T \rangle$,
the value of the standard
deviation $\delta T=\sqrt{\langle T^2 \rangle - \langle T \rangle^2}$  for   systems with $\alpha=1/3$ is  larger than  those  with  $\alpha=1/2$
(see the caption in Fig. 3), although, for the particular
cases shown in Fig. \ref{fig_3}(a) and  Fig. \ref{fig_3}(c), as well as in Fig. \ref{fig_3}(b) and  Fig. \ref{fig_3}(d), the  difference between the  standard deviations  is small.

\begin{figure}
\includegraphics[width=\columnwidth]{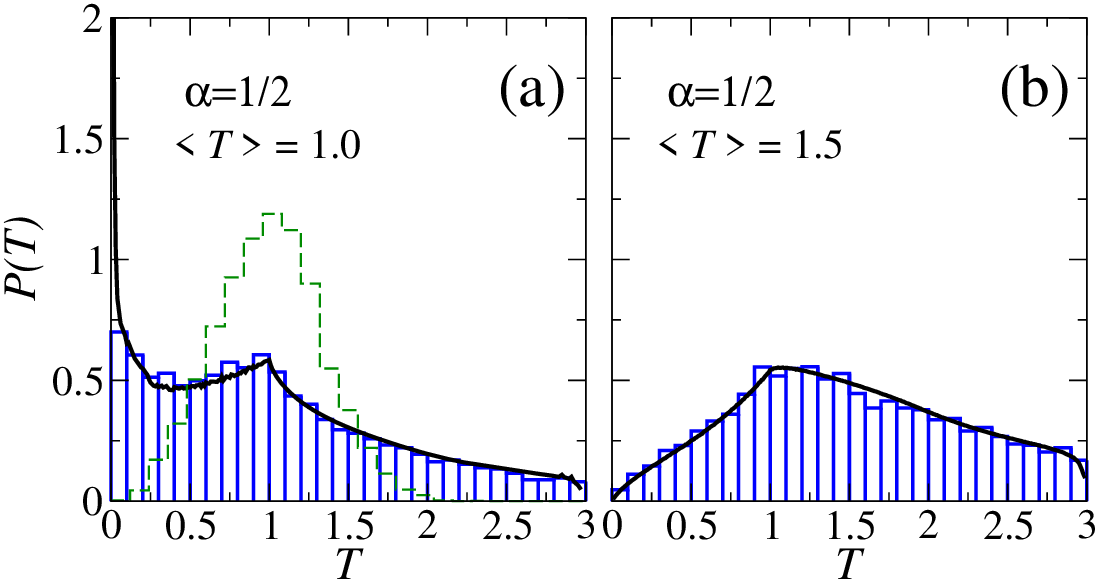}
\includegraphics[width=\columnwidth]{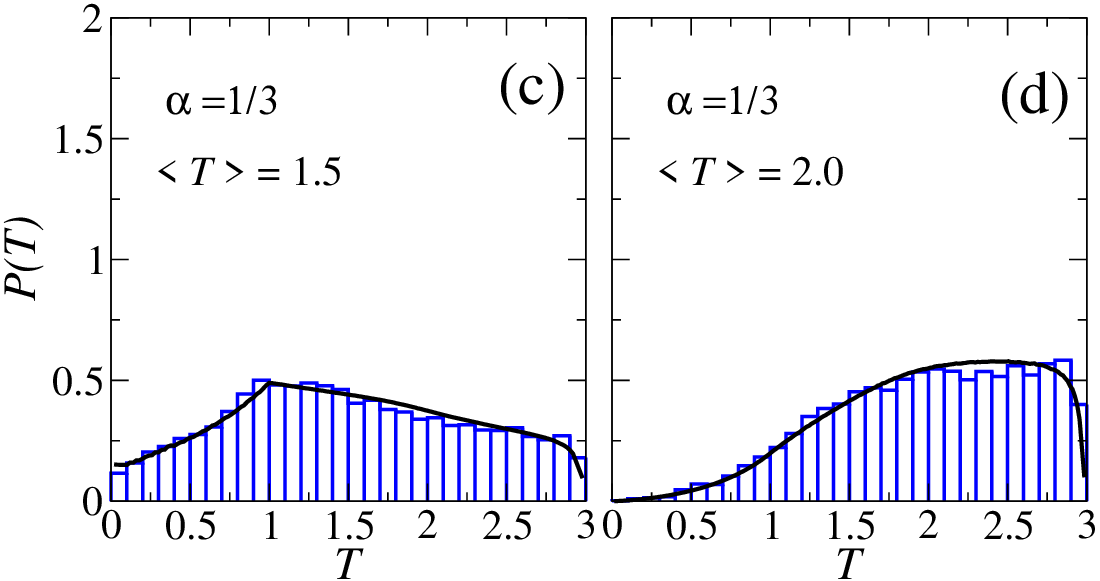}
\caption{(Color online)  Transmission distribution $P(T)$ for $N=3$ transmission channels with L\'evy disorder characterized by  $\alpha=1/2$ and 1/3, upper and lower
panels, respectively. The solid lines are obtained according to the theory described in the main text, while the histograms are extracted from the tight binding numerical simulations.
The values of the standard deviation $\delta T$ and the constant $b$ in $s(\alpha,\mu,N, z)$ for each panel are: (a) $\delta T=0.73$, $b=2.8$, (b) $\delta T=0.70$, $b=1.36$,
(c) $\delta T=0.75$, $b=2.6$, (d) $\delta T=0.62$,  $b=0.5$. A good agreement between theory and numerical simulation can be seen in all panels. (a) The histogram
(green dashed-line) shows the distribution $P(T)$ for systems with standard disorder with ensemble average $\langle T \rangle =1.0$.  }
\label{fig_4}
\end{figure}

We now consider the case of $N=3$ transmission channels. Thus,  the maximum value of the transmission is  3.  In Fig. \ref{fig_4}, we show
the transmission distribution for $\alpha =1/2$
and $\alpha=1/3$ at different transmission averages $\langle T \rangle$. The  solid lines are calculated as given by Eq. (\ref{pofG_mu}), while the histograms are
obtained from the numerical simulations.
As in the two-transmission channel case,  disorder configurations with $\alpha=1/3$
show larger
transmission fluctuations  than  $\alpha=1/2$. See for instance the distributions in Fig. \ref{fig_4} (b) and
(c), which have the same average $\langle T \rangle=1.5$, but  $\delta T =0.7$ and 0.75, respectively.
In all the  panels of  Fig. \ref{fig_4} a good agreement between theory (solid lines) and numerical simulations
(histograms) can be seen.

Finally, we remark that the landscape  of the transmission  distributions for  standard (dashed-line histogram) and L\'evy disordered (solid line) systems shown in
Fig. \ref{fig_3}(a), as well in  Fig. \ref{fig_4}(a),
are quite different.  In general, the transmission distributions in the presence of L\'evy disorder are wider than
than the cases in the presence  of standard Anderson localization, revealing stronger  transmission fluctuations  in the  former case.

\subsubsection{Broken time-reversal symmetry}

We first recall  that under the presence of time-reversal symmetry,  the reflection probability is slightly higher than the
transmission probability due to constructive interference between two  time-reversed scattering processes. This phenomenon is known as weak localization.
If time-reversal symmetry is broken this constructive interference effect is destroyed and the weak localization is suppressed. Therefore, it is expected that the  absence of
time-reversal symmetry has an effect on the statistics of the transmission.

Let us assume now that we break the time-reversal symmetry of the L\'evy disordered systems, i.e., we consider the symmetry class
$\beta=2$ . In the numerical simulations, time-reversal symmetry is broken by applying a perpendicular magnetic field to the disordered systems.

\begin{figure}
\includegraphics[width=\columnwidth]{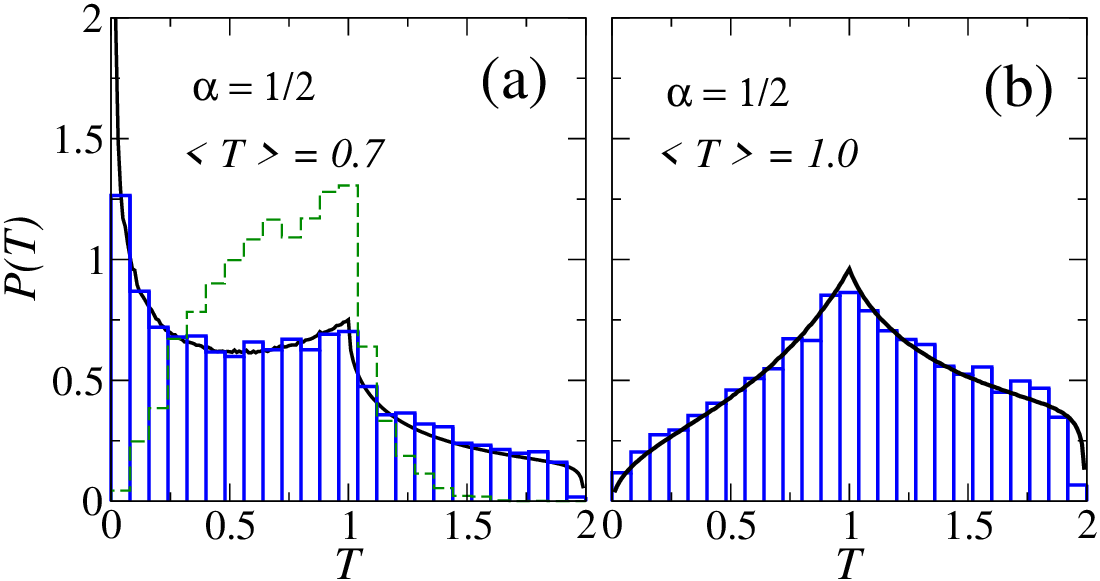}
\includegraphics[width=\columnwidth]{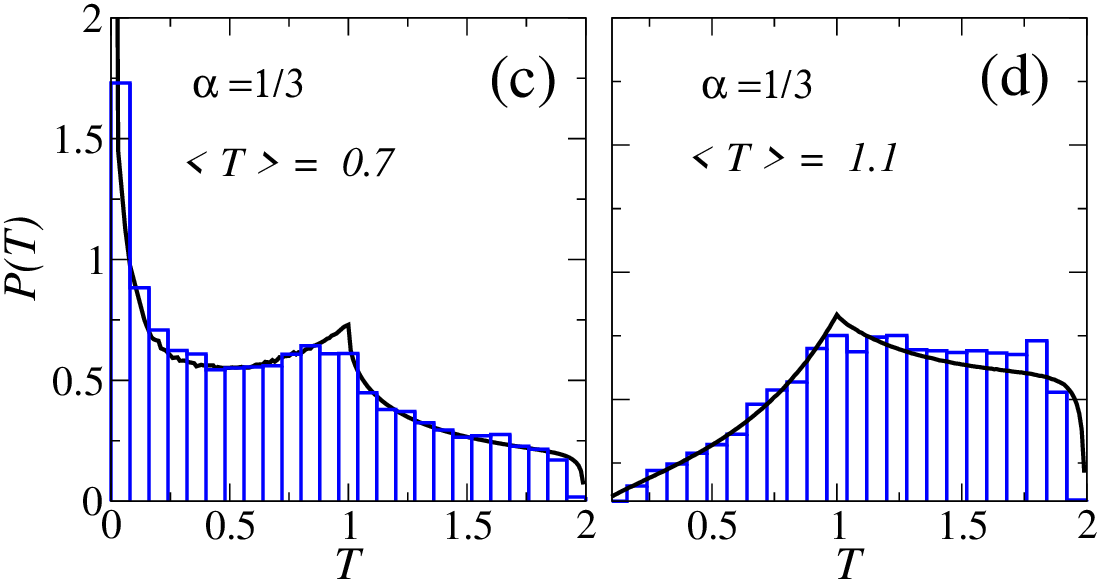}
\caption{(Color online) Applied  magnetic field ($\phi=0.15$).  Transmission distribution $P(T)$ with broken time-reversal symmetry   for $N=2$ transmission channels with L\'evy disorder characterized by  $\alpha=1/2$ and 1/3, upper and lower
panels, respectively. The solid lines are obtained according to the theory described in the main text, while the histograms are extracted from the tight binding numerical simulations.
The values of the standard deviation $\delta T$ and the constant $b$ in $s(\alpha,\mu,N, z)$ for each panel are: (a) $\delta T=0.5$, $b=3.3$,  (b) $\delta T=0.46$, $b=1.3$,
(c) $\delta T=0.52$, $ b=2.5$ and (d) $\delta T=0.46$, $b=0.8$. Theory and numerical simulations are in agreement in all panels. (a) The histogram
(green dashed-line) shows the distribution $P(T)$ for systems with standard disorder and broken time-reversal symmetry  with ensemble average $\langle T \rangle =0.7$. }
\label{fig_5}
\end{figure}
\begin{figure}
\includegraphics[width=\columnwidth]{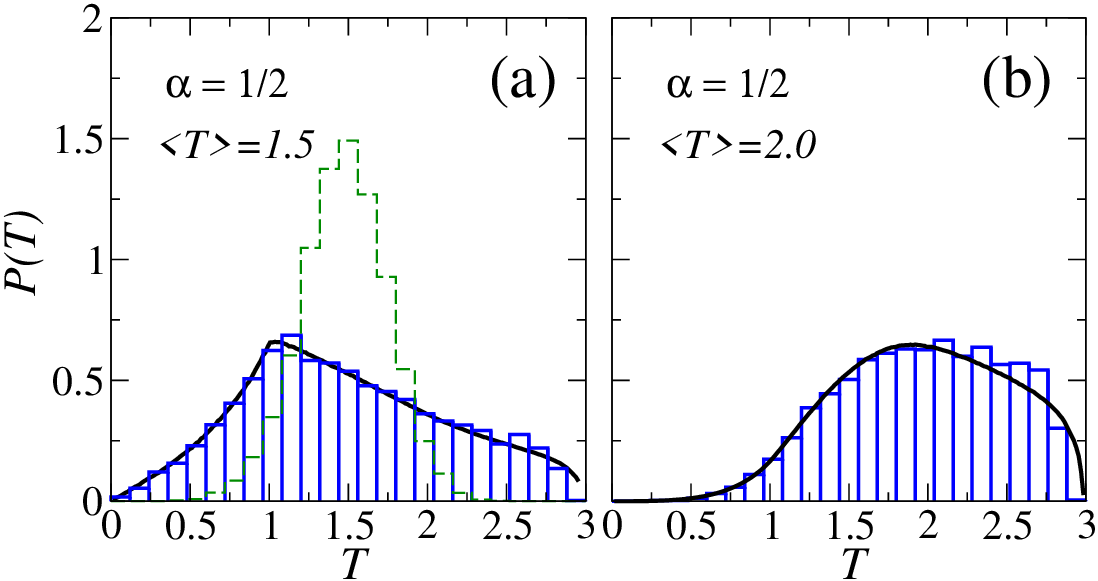}
\includegraphics[width=\columnwidth]{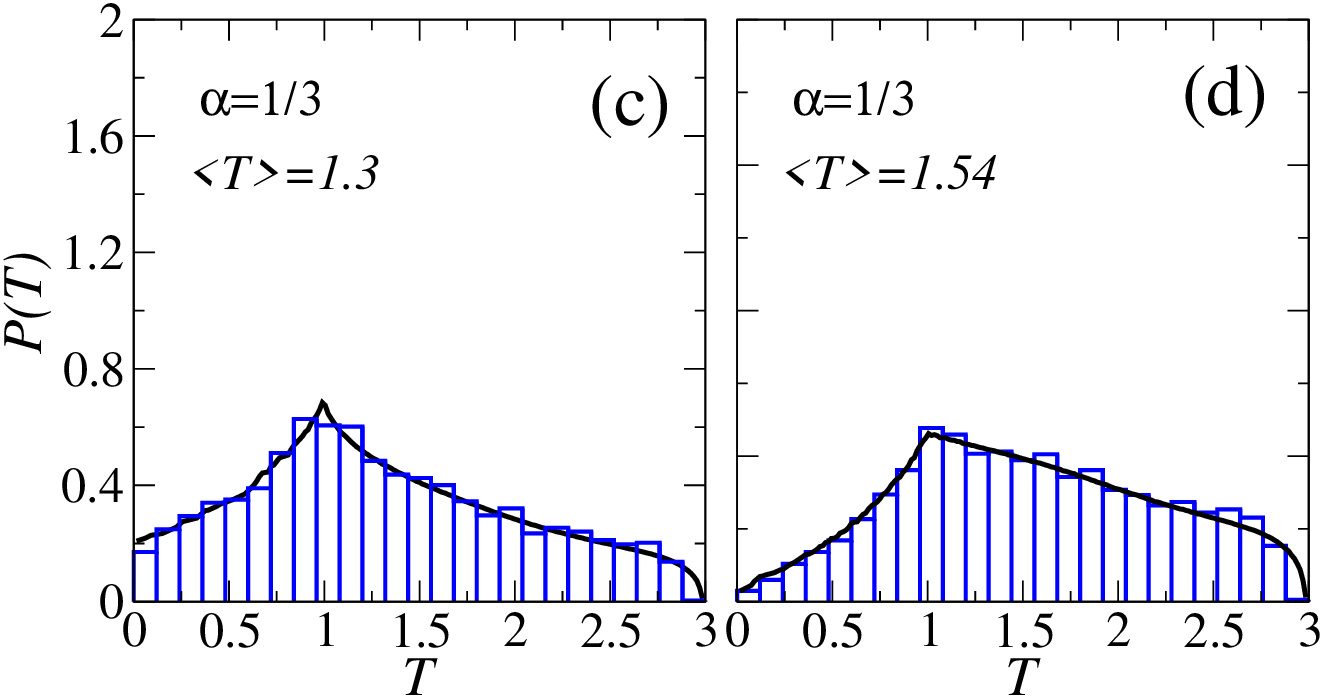}
\caption{(Color online) Applied magnetic field ($\phi=0.1$). Transmission distribution $P(T)$ with broken time-reversal symmetry   for $N=3$ transmission channels with
L\'evy disorder characterized by  $\alpha=1/2$ and 1/3, upper and lower
panels, respectively. The solid lines are obtained according to the theory described in the main text, while the histograms are extracted from the tight binding numerical simulations.
The values of the standard deviation $\delta T$ and the constant $b$ in $s(\alpha,\mu,N, z)$ for each panel are: (a) $\delta T=0.50$,  $b=1.6$,  (b) $\delta T=0.51$, $b=0.7$,
(c) $\delta T=0.70$, $b=1.9$ and  (d) $\delta T=0.67$, $b=1.6$. Theory and numerical simulations are in agreement in all panels.  (a) The histogram
(green dashed-line) shows the distribution $P(T)$ for systems with standard disorder and broken time-reversal symmetry  with ensemble average $\langle T \rangle =1.5$.}
\label{fig_6}
\end{figure}

In Fig. \ref{fig_5} we show  the transmission distribution for $N=2$  channels with $\alpha =1/2$ and 1/3, Figs. \ref{fig_5}(a)-(b) and
\ref{fig_5}(c)-(d), respectively. Disordered systems  with approximately the same average $\langle T \rangle$ were chosen in
Figs. \ref{fig_5}(a) and \ref{fig_5}(c) [as well as Figs. \ref{fig_5}(b) and \ref{fig_5}(d)] for their comparison.  Similarly,   in  Figs. \ref{fig_6}(a)-(b) and Figs. \ref{fig_6}(c)-(d) we show
the transmission distributions for  $N=3$ channels with $\alpha =1/2$ and 1/3, respectively .

As in  the case of preserved time-reversal symmetry in the previous subsection, smaller values of $\alpha$, i.e., a larger tail of the L\'evy distribution,  lead to stronger
transmission fluctuations $\delta T$, for a fixed value of the average $\langle T \rangle$. For instance, Figs. \ref{fig_5}(a)  and \ref{fig_5}(c) show a couple of distributions  $P(T)$
both with $\langle T \rangle=0.7 $, but   $\delta T=0.5$ and 0.52 for $\alpha=1/2$ and 1/3, respectively.

It is also interesting to compare the transmission distributions in absence of time-reversal symmetry (Figs. \ref{fig_5} and \ref{fig_6}) with those previously shown for the
case of preserved  time-reversal symmetry (Figs. \ref{fig_3} and \ref{fig_4}). As we have mentioned, when time-reversal symmetry is present,
constructive
interference leads to an enhancement of the reflection. In general, this enhancement is small, but
one can observe its effects at the level of the distribution $P(T)$: for instance, at small transmission values (or high reflection),  Fig. \ref{fig_3} (b) shows that the
transmission probability is larger compared  to
the broken time-reversal symmetry in Fig. \ref{fig_5} (b), i.e.,  reflection is enhanced. This enhancement in
the reflection  is perhaps better seen  by comparing the distributions in Figs. \ref{fig_3} (d) and \ref{fig_5} (d),  although in
these Figs. the average $\langle T \rangle$ is not exactly the same; we can observe that  $P(T)$ in Fig. \ref{fig_5} (d)  is suppressed in absence of time-reversal at small values of $T$
and therefore it is  less symmetric in respect to $T=1$ than $P(T)$  in Fig. \ref{fig_3} (d), i.e., reflection processes are promoted when time-reversal
symmetry is present.

Finally, we remark  the strong effect of the presence of L\'evy disorder in relation to
standard disorder systems.  In Figs. \ref{fig_5}(a) and \ref{fig_6}(a) we have included (green-dashed line histograms)
the transmission distributions for disordered systems with standard Anderson  localization, which, as we can see,  have a complete different landscape than  those of the
L\'evy disordered  systems.  In general, the transmission fluctuations are larger in the presence of L\'evy disorder and therefore the transmission
distributions are wider than  in the presence of  standard Anderson localization.

\section{Summary and Conclusions} \label{conclusions_section}

Most of the research on transport of classical  and quantum waves, such as electromagnetic fields and  electrons, through random media uses
distributions with finite moments to  model the disorder in the media. Using these standard disorder models  several
theoretical approaches have studied  properties
of the wave transport, such as the widely known phenomenon of  Anderson localization. In particular, a scaling theory of localization has been developed to study the statistical
properties of the transport through  disordered systems. Within that framework and using random-matrix theory, it has been shown that for one-dimensional
and quasi-dimensional disordered systems, a single parameter, the localization length, determines the statistical properties of the transmission.

On the other hand,  there is a family of probability density functions (L\'evy distributions) whose first  moment diverges due to their long tails, which are   characterized by  the
exponent  $\alpha$ of the power-law tail.  L\'evy distributions  emerge in several and very different phenomena and  different areas,  such as economy and biology.

In the past  [\onlinecite{Falceto}], we have introduced those heavy-tailed distributions to model
disorder  in random media and study their effects on  the transport, however, we restricted ourselves to the case of a single transmission channel. It
was found that the statistical properties of the dimensionless conductance (transmission) are completely determined by two parameters:
the localization length and the power $\alpha$. It was also found that waves become less localized, or anomalously localized,
in relation to the case of Anderson localization.

The present work is a generalization of the previous study in Ref. [\onlinecite{Falceto}] to consider L\'evy disordered systems whose total transmission is given by the  contribution of
several  channels. This is also of experimental relevance since it imposes  less
restrictive conditions than considering systems with  a single transmission channel.

Thus, by extending the scaling approach to localization for multichannel standard disordered systems, we have calculated the transmission distribution for multichannel
L\'evy disordered systems, which is determined by the power $\alpha$ and the average $\langle \ln T \rangle$. We  show several examples  of the transmission distribution  for systems with  2 and 3 transmission
channels. The theoretical results  have been verified by  tight binding numerical simulations.  Additionally, we have studied the effects of breaking  time-reversal
symmetry in the L\'evy disordered  systems.

We have contrasted the transmission distributions for L\'evy and standard disordered systems and showed that the landscape of both distributions is very different.
In general, the transmission distributions for  L\'evy disordered systems  are wider due to
the strong random fluctuations of the transmission
than those obtained for standard disordered systems

Finally, we have confirmed all our theoretical results by comparison  with  tight binding numerical simulations. Nevertheless, it would be highly desirable to
verify experimentally the effects of L\'evy disorder on the transport  like those we have studied here. For instance, L\'evy disorder may be implemented in
random microwave-waveguides and/or random optical-fibers  experimental setups.

$^\dagger$Current affiliation: Physics Division, National Center for Theoretical Sciences, Hsinchu 300, Taiwan

\acknowledgments
We would like to thank Xujun Ma for bringing Ref. [\onlinecite{Engin}] to our attention.
V. A. G  acknowledges support from MINECO (Spain) under the Project number FIS2015-65078-C2-2-P and Subprograma Estatal de Movilidad 2013-2016 under
the Project number PRX16/00166. He also thanks for the hospitality the  Physics Department of Queens College, The City University of New York, where part of this work was written.
F. F.  acknowledges support from MINECO (Spain) under the Project number FPA2015-65745-P and DGA 2014-E24/2.
The authors thankfully acknowledge the resources from the supercomputer ``Terminus'', technical expertise and
assistance provided by the Institute for Biocomputation and Physics of Complex Systems (BIFI) - Universidad de Zaragoza.
Ilias Amanatidis acknowledges support provided by the
Center for Theoretical Physics of Complex Systems in Daejeon of Korea under the project
IBS-R024-D1. We  also acknowledge support from the National Taiwan University and
the Ministry of Science and Technology of R.O.C. Taiwan.

\begin{appendix}
\subsection{Appendix. Numerical simulations}

 In this appendix,  we present the numerical model that was used to verify the theoretical predictions of the previous section. We consider a standard single-orbital tight-binding square lattice with Hamiltonian
\begin{equation}
 H=\sum_i \epsilon_i c_i^\dagger c_i +  \sum_{<ij>}(t_{ij}c_i^\dagger c_j +\text{h.c.}),
\label{h_square}
\end{equation}
where $\epsilon_i$ is the on-site energy at site $i$, while $t_{ij}$ represents the nearest neighbor hopping between sites $<ij>$ and $c_{i}^\dagger(c_{i})$ is the corresponding
creation(annihilation) operator for electrons. For simplicity we set $t_{ij}=t=1$
and the lattice constant to 1. In this model, the disorder is implemented by random on-site energies $\epsilon_i$, sampled  from a uniform
distribution in the interval $[-w/2,w/2]$.
Along this paper, the statistics of the transmission probability are collected from 10000 different disorder realizations.
In order to make the numerical model statistically equivalent to the theoretical model,
we consider that the length of the square lattice at each disorder realization is determined
by the number of scatterers, whose intermediate spacings are sampled from the L\'evy distribution,
that can be fitted in a system of length $L$ in the theoretical model.

The transmission probability can be calculated by attaching perfect leads from left and right,
described by Eq. \ref{h_square} for $\epsilon_i=0$ and then applying the Green's function method [\onlinecite{Datta}].
The Green's function is given by,
\begin{equation}
      G(E) = (EI - H  - \Sigma_{L}(E) -  \Sigma_{R}(E) )^{-1}
\end{equation}
where $\Sigma_{L(R)}(E)$ is the self-energies of the left(right)
lead and E is the incident energy of the electrons. The self-energies follow a matrix form
\begin{equation}
      \Sigma_{L(R)}(E;n,m) =  \sum_{j=1}^{M} \chi_{j}(n)g(E,j)\chi_{j}(m)
\end{equation}
where $M$ is the number of sites
transverse to the transport direction where hard wall boundary conditions are applied and $j$ is
an integer taking values $j=1,2..M$.
We fix the energy at $E=0.1t$ so that $M$ determines the number of open transmission channels.

The surface Green's function of the square lattice leads $g(E,j)$ at site $j$ is given by [\onlinecite{Lewenkopf}],
\begin{equation}
  g(E,j) = \frac{(E-\epsilon(j))}{2} -i \sqrt{1-\frac{(E-\epsilon(j))^{2}}{4}}
\end{equation}
with  $\epsilon(j) = 2\cos(\frac{\pi j }{M+1}) $ and $|E-\epsilon(j)|<2$, while
$\chi_{j}(n)$ are the transverse wavefunctions due to the hard-wall boundary conditions,

\begin{equation}
\chi_{j}(n) =  \sqrt{\frac{2}{M+1}} \sin\left(\frac{\pi jn}{M+1}\right)
\end{equation}
with  $n=1,...M$. Then, the transmission probability can be calculated by [\onlinecite{Datta}],

\begin{equation}
      T(E) = Tr[\Gamma_{L}(E)G(E)\Gamma_{R}(E)G(E)^{\dagger }]
\end{equation}
where the matrices $\Gamma_{L}(E),\Gamma_{R}(E)$ are related with the velocities of the incident electrons
and can be calculated calculated via the self-energies from,
\begin{equation}
    \Gamma_{L(R)}(E) =  i\Big[\Sigma_{L(R)}  - \Sigma_{L(R)}^{\dagger } \Big].
\end{equation}

Finally, in the case that we break the time-reversal symmetry of the disordered system by applying
a  magnetic field transverse to the plane of the 2D wire, the tight-binding Hamiltonian for our numerical simulations is given by Eq. \ref{h_square}, with a modified hopping $t_{ij}$,

\begin{equation}
        t_{ij} = e^{i\phi_{ij}}.
\end{equation}
The factor $\phi_{ij}$ is the Peierls phase(see Ref. [\onlinecite{Lewenkopf}]) between sites $i$ and $j$ given by,

\begin{equation}
        \phi_{ij} = \frac{2\pi}{\Phi_{0}}\int_{r_{i}}^{r_{j}}\textbf{A}d\textbf{l}
\end{equation}
where $\Phi_{0}$ is the flux quantum defined as $\Phi_{0}=h/ce$.
We assume that the vector potential $\textbf{A}$ is along
the transport direction x, that is $\textbf{A}=-By\hat{x}$, corresponding to a homogeneous
out of plane magnetic field $\textbf{B}=B\hat{z}$. The phase factor
$\phi_{ij}$ then becomes,

\begin{equation}
        \phi_{ij} = \frac{2\pi B }{\Phi_{0}}(x_{j}-x_{i})\Big( \frac{y_{j}+y_{i}}{2} \Big)
\end{equation}
which is non-zero only for the horizontal hoppings in the square lattice.
In all the numerical simulations we measure the magnetic field strength via the flux per square
plaquette $\Phi=Ba^{2}$ in the square lattice, in units of $\Phi_{0}$.

\end{appendix}


\begin{thebibliography}{50}

\bibitem{Anderson}
P. W.  Anderson, Phys. Rev. {\bf 109}, 1492 (1958);

\bibitem{50Anderson}
A. Lagendijk, B. van Tiggelen, and D. S. Wiersma, Physics Today {\bf  62}(8), 24 (2009).

\bibitem{Fifty_years}
{\it Fifty years of Anderson  localization}, edited by E. Abrahams (World Scientific, Singapore, 2010).

\bibitem{Bose-Einstein}
G. Roati, C.  D'Errico, L. Fallani, M.  Fattori, C. Fort, M. Zaccanti, G. Modugno, M.  Modugno, M. Inguscio,  Nature, {\bf 453}, 895, (2008).

\bibitem{Anderson_photons}
A.  Crespi, R. Osellame, R.  Ramponi, V. Giovannetti, R. Fazio, L.  Sansoni, F.  De Nicola, F. Sciarrino,  P. Mataloni, Nature Photonics {\bf 7},
322  (2013).

\bibitem{Chabanov}
A. A. Chabanov, M. Stoytchev, and A. Z. Genack, Nature
(London) {\bf 404}, 850 (2000).

\bibitem{Shi}
Z. Shi, J. Wang, and A. Z. Genack, Proc. Natl. Acad. Sci.
U.S.A.  {\bf 111}, 2926 (2014).

\bibitem{Penha}
A. Peña, A. Girschik, F. Libisch, S. Rotter, and A. A.
Chabanov, Nat. Commun. {\bf 5}, 3488 (2014).

\bibitem{Yamilov}
 A. G. Yamilov, R. Sarma, B. Redding, B. Payne, H. Noh,
and H. Cao, Phys. Rev. Lett. {\bf 112}, 023904 (2014)


\bibitem{Mello_book}
P. A. Mello and N. Kumar, {\it Quantum Transport in Mesoscopic Systems: Complexity and Statistical Fluctuations} (Oxford
University Press, Oxford, 2004).

\bibitem{Beenakker-review}
C. W. J. Beenakker, Rev. Mod. Phys. \textbf{69}, 731 (1997).

\bibitem{Soukoulis}
C. M. Soukoulis and E. N. Economou, Phys. Rev. B {\bf 24}, 5698 (1981).

\bibitem{Evangelou}
S. N. Evangelou and D. E. Katsanos, J. Phys. A: Math Gen. {\bf 36}, 3237 (2003).

\bibitem{We} I. Amanatidis, I. Kleftogiannis, F. Falceto, and V. A. Gopar
Phys. Rev. B {\bf 85}, 235450 (2012).

\bibitem{nanoribbons}
I. Kleftogiannis, I. Amanatidis, and V. A. Gopar,  Phys. Rev. B  {\bf 88},  205414  (2013).

\bibitem{Barthelemy}
P.  Barthelemy, J. Bertolotti  and D. S. Wiersma, Nature,  {\bf 453}, 495  (2008).

\bibitem{Antonio_prl}
A. A. Fern\'andez-Mar\'in, J. A. M\'endez-Berm\'udez, J. Carbonell, F. Cervera, J. S\'anchez-Dehesa, and V. A. Gopar
Phys. Rev. Lett. {\bf 113}, 233901 (2014).

\bibitem{Michael}
J. Klafter, M. F Shlesinger, G. Zumofen, Physics Today {\bf 49}(2), 33 (1996).

\bibitem{stock}
Mantegna R. N. \and Stanley H. E.,  Nature {\bf 376}, 46 (1995).

\bibitem{Lambert}
M. Leadbeater, V. I. Falko, C. J.  Lambert, Phys. Rev. Lett.{\bf 81} 1274 (1998).

\bibitem{Hideo}
H. Kohno and H. Yoshida, Solid State Commun, {\bf 132}, 59 (2004).

\bibitem{fishes}
Sims D. W., et. al. Nature,{\bf 45}, 1098 (2008).

\bibitem{Mercadier}
N. Mercadier, W. Guerin, M. Chevrollier, and R. Kaiser, Nature Phys. {\bf 5}, 602,  (2009).

\bibitem{Levy}
P. L\'evy, {\it Th\'eorie de l'addition des variables al\'eatoires}, (Gauthiers-Villars, Paris, 1937).

\bibitem{Kolmogorov}
B. V. Gnedenko and A. N. Kolmogorov, {\it Limit distributions for sums of independent random variables},
(Addison-Wesley,  Cambridge, MA, 1954).

\bibitem{Uchaikin}
V. V. Uchaikin and V. M. Zolotarev, {\it Chance and Stability. Stable Distributions and their Applications}
(VSP,  Utrecht, Netherlands, 1999) and references therein.

\bibitem{largervalues} One  can extend the present model to the case $1 < \alpha <2$, however,
we restrict ourselves to  $0 < \alpha <1$, where effects of anomalous localization are stronger.

\bibitem{Falceto}
F.  Falceto and Victor A. Gopar, EPL \textbf{92}, 57014 (2010).

\bibitem{incoherent}
For 1D systems with  L\'evy disorder in the incoherent regime,  see for instance  Refs. [\onlinecite{Boose,Burioni, Beenakker,Sibatov}].

\bibitem{Anderson_1}
P. W. Anderson, D. J. Thouless, E. Abrahams E. and D. S. Fisher, Phys. Rev. B \textbf{22}, 3519,(1980).

\bibitem{Melnikov}
V. I. Mel’nikov, Pis’ma Zh. Eksp. Teor. Fiz. {\bf 32}, 244 (1980); [JETP Lett. {\bf 32}, 225 (1980)].

\bibitem{qalpha_fourier} The explicit expression for the probability density ${q}_{\alpha,c}(x)$ is more conviniently written using the Fourier transform:
$\widehat {q}_{\alpha,c}(k)= \exp \left(-|k|^\alpha\left( A\theta(k)+
A^*\theta(-k) \right)\right) $,  where $\theta$ is the Heaviside step function and $A= -c
{\Gamma(-\alpha)} {\rm e}^{i\frac{\pi\alpha}2}$, $\Gamma$ being the
Gamma function.

\bibitem{Abrikosov}
A. A. Abrikosov, Solid State Commun. {\bf 37}, 997 (1981).

\bibitem{Mello_groups}
P. A. Mello, J. Math. Phys. \textbf{27}, 2876 (1986).

\bibitem{Engin} E. E. Kuruo\v{g}lu, IEEE transactions on signal processing {\bf 49}, 2192 (2001).

\bibitem{dmpk}
P. A. Mello, P. Pereyra, and N. Kummar, Ann. Phys. (N. Y.)  \textbf{181}, 290 (1988).


\bibitem{Boose}
D. Boos\'e and J. M. Luck, J Phys. A. Theor. {\bf 40}, 140405, (2007).

\bibitem{Burioni}
R. Burioni, L. Caniparoli, and A. Vezzani, Phys. Rev. E {\bf 81}, 060101(R) (2010).

\bibitem{Beenakker}
C. W. J. Beenakker, C. W. Groth, and A. R. Akhmerov, Phys. Rev. B, {\bf 79}, 024204, (2009).

\bibitem{Sibatov}
R. T. Sibatov, JETP Letters, {\bf 93},  503 (2011).


\bibitem{Rejaei}
C. W. J. Beenakker, and B. Rejaei, Phys. Rev. Lett. \textbf{71}, 3689 (1993).

\bibitem{Muttalib}
K. A. Muttalib, P. W\"{o}lfle, V. A. Gopar, Annals of Physics {\bf 308}, 156 (2003).

\bibitem{crossover}
V.  A.  Gopar, K. A. Muttalib, P. W\"olfle, Phys. Rev. B, {\bf 66}, 174204 (2002).

\bibitem{Datta}
S. Datta, {\it Electronic Transport in Mesoscopic Systems},
(Cambridge University Press, 1997).

\bibitem{Lewenkopf}
C. H. Lewenkopf, E. R.   Mucciolo,  J. Comput. Eletron., {\bf 12}, 203 (2013).
 	
\end{thebibliography}
\end{document}